\documentclass[aps,letterpaper,twocolumn,pra,footinbib,floatfix,showpacs]{revtex4}
\usepackage{amsmath,verbatim,graphicx,natbib,adjustbox}

\begin{document}
\author{Bhuvanesh Sundar}
\affiliation{Laboratory of Atomic and Solid State Physics, Cornell University, Ithaca New York 14850, USA}
\author{Erich J. Mueller}
\affiliation{Laboratory of Atomic and Solid State Physics, Cornell University, Ithaca New York 14850, USA}
\title{Lattice bosons with infinite-range checkerboard interactions}
\date{\today}

\begin{abstract}
Motivated by experiments performed by Landig \textit{et al.} [Nature (London) {\bf 532}, 476 (2016)], we consider a two-dimensional Bose gas in an optical lattice, trapped inside a single mode superradiant Fabry-Perot cavity. The cavity mediates infinite-range checkerboard interactions between the atoms, which produces competition between Mott insulator, charge-density wave, superfluid, and supersolid phases. We calculate the phase diagram of this Bose gas in a homogeneous system and in the presence of a harmonic trap.
\end{abstract}

\newcommand{\hc}{\hat{c}}
\newcommand{\+}{^\dagger}
\newcommand{\ket}[1] {\left| {#1} \right\rangle}
\newcommand{\expect}[1] {\left\langle {#1} \right\rangle}

\maketitle

\section{Introduction}\label{sec:intro}

Introducing long-range interactions between bosonic atoms in an optical lattice provides the opportunity to explore different phases, driven by the competition between short-range interactions, long-range interactions, and quantum tunneling. Interactions mediated via an optical cavity provide an avenue to explore this physics \cite{Esslinger,hemmerich}. In this paper, we calculate the phase diagram of bosonic atoms experiencing such cavity-mediated long-range interactions.
We find a rich phase diagram with superfluid (SF), supersolid (SS), Mott insulator (MI), and charge-density wave (CDW) phases. We find a breakdown of the local-density approximation (LDA), and good agreement with experiments \cite{Esslinger}.

By trapping $^{87}$Rb atoms in a transversely pumped single mode optical cavity and tuning the cavity into the superradiant phase, Landig \textit{et al.} have produced infinite-range interactions between bosonic atoms \cite{Esslinger}. Interference between the pump beam and the light scattered into the cavity results in a checkerboard intensity pattern, whose strength is proportional to the number of atoms on the high intensity sites. Integrating out the photons yields a long-range checkerboard interaction. By changing the lattice depth and the cavity detuning, the experimentalists can independently tune the strengths of the short-range and long-range atomic interactions relative to the tunneling strength. Adding these long-range interactions to a Bose-Hubbard model, we use a variational ansatz to produce a phase diagram. We consider both a homogeneous and a harmonically trapped system. In addition to SF order, characterized by off-diagonal long-range order in the single-particle density matrix, the system can also display CDW order, where the occupations on the even and odd sites differ. Coexistence of both orders results in a SS, and the absence of both orders a MI. All four of these phases are found in our calculations, and were seen in experiments as well \cite{Esslinger}. We predict that a reanalysis of existing experimental data will reveal previously undetected phase transitions. Some of our results for homogeneous systems have been seen in other theoretical studies \cite{Donner, HuiZhai,Hofstetter,Panas}.

This paper is organized as follows. In Sec. \ref{sec:homogeneous}, we introduce our model for a homogeneous Bose gas, and present the phase diagram. In Sec. \ref{sec:inhomogeneous}, we analyze the harmonically trapped case. We conclude in Sec. \ref{sec:conclusions}.

\section{Homogeneous gas}\label{sec:homogeneous}

In this section, we explore the phase diagram of a homogeneous Bose gas in an optical lattice, trapped inside a single mode optical cavity. We calculate the ground state of the bosons by minimizing the energy of a variational many-body wave function. We obtain the phase boundaries through a combination of numerical and analytical means. This section is organized as follows. In Sec. \ref{subsec:model1}, we introduce our model for a homogeneous Bose gas in an optical lattice, including cavity-mediated infinite-range interactions. In Sec. \ref{subsec:ansatz1}, we introduce our variational ansatz and calculate the energy of the system. In Sec. \ref{subsec:phaseDiag1}, we present the phase diagram of our model.

\subsection{Model}\label{subsec:model1}

\begin{figure}[t]
\begin{center}
\includegraphics[width=0.8\columnwidth]{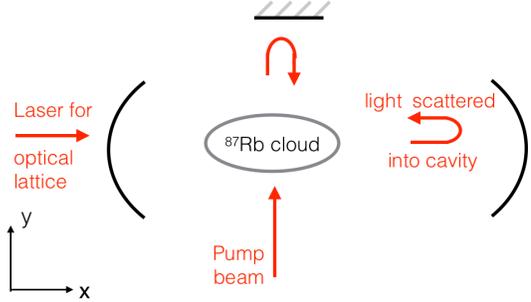}
\end{center}
\caption{Schematic of the experimental setup. A trapped cloud of bosonic $^{87}$Rb atoms sits in an off-resonant optical cavity. The atoms are pumped from the side, and scatter light into the cavity. A second laser enters the cavity along its axis, producing an optical lattice.}
\label{fig:schematic}
\end{figure}

We consider an ultracold gas of bosons tightly confined to the $x$-$y$ plane, where they experience a 2D optical lattice. The atoms are coupled to the fundamental mode of a high finesse Fabry-Perot cavity oriented along the $x$ direction, and illuminated by a pump beam along the $y$ direction (see Fig. \ref{fig:schematic}). The pump light scattered by the atoms into the cavity mediates an effective infinite-range atom-atom interaction. The effective atom-atom interactions are derived in \cite{RitschModel}, producing a Hamiltonian
\begin{equation}\label{eqn:model1}
 \hat{H} = \hat{H}_{\rm lat} + \hat{H}_{\rm cav}.
\end{equation}
The term $\hat{H}_{\rm lat}$ models the trap, tunneling of atoms, and on-site interactions in the two-dimensional optical lattice:
\begin{equation}\label{eqn:Hlat}
 \hat{H}_{\rm lat} = \sum_{\langle ij\rangle} -J\hc_i\+\hc_j + \rm{H.c.} + \sum_i \frac{U}{2}\hc_i\+\hc_i\+\hc_i\hc_i - \mu_i\hc_i\+\hc_i.
\end{equation}
The operator $\hc_i\+ (\hc_i)$ creates (annihilates) a boson at lattice site $i$. The hopping strength $J$ can be tuned by controlling the intensity of the laser creating the optical lattice. The on-site interaction strength $U$ can be controlled via the laser intensity and the transverse confinement, or by tuning the magnetic field near a Feshbach resonance. The last term in Eq. (\ref{eqn:Hlat}) models the trap, where $\mu_i$ is an effective spatially dependent chemical potential,
\begin{equation}
 \mu_i = \mu - \frac{1}{2}m\omega^2\left(x_i^2+y_i^2\right),
\end{equation}
where $x_i$ and $y_i$ denote the co-ordinates of lattice site $i$ in integer multiples of the lattice constant. For a homogeneous gas, we set $\omega=0$.

The term $\hat{H}_{\rm cav}$ models the infinite-range interactions mediated by the light in the Fabry-Perot cavity,
\begin{equation}\label{eqn:Hcav}
 \hat{H}_{\rm cav} = -\frac{U'}{K}\left(\sum_i (-1)^{x_i+y_i}\hc_i\+\hc_i\right)^2,
\end{equation}
where $K$ is the total number of lattice sites. The effective long-range interaction strength $U'$ is related to experimental parameters as
\begin{equation}
 U' \simeq -K\frac{\hbar\eta^2}{\Delta_c}
\end{equation}
where $\eta$ is the two-photon Rabi frequency, and $\Delta_c$ is the detuning of the optical lattice laser from the fundamental mode of the Fabry-Perot cavity \cite{RitschModel}. In this paper, we only work in a regime where $U'>0$. When at fixed density, the long-range interaction energy scales as $K^2$, while all other energies scale as $K$. Thus, to achieve a reasonable thermodynamic limit, we fix $U'$ while $K\rightarrow\infty$.

\subsection{Gutzwiller ansatz}\label{subsec:ansatz1}

The model in Eq. (\ref{eqn:model1}) breaks the symmetry between two kinds of sites: those for which $x_i+y_i$ is even (which we call even sites), and for which $x_i+y_i$ is odd (which we call odd sites). We make a variational ansatz which includes this asymmetry:
\begin{equation}\label{eqn:Gutzwiller1}
 \ket{\psi} = \left(\sum_{i\in {\rm even}}\sum_{n=0}^\infty \frac{a_n}{\sqrt{n!}}\left(\hc_i\+\right)^n\right) \left(\sum_{j\in {\rm odd}}\sum_{n=0}^\infty \frac{b_n}{\sqrt{n!}}\left(\hc_j\+\right)^n\right)\ket{0},
\end{equation}
where $\ket{0}$ is the vacuum of atoms. Our ansatz in Eq. (\ref{eqn:Gutzwiller1}) is an extension of the Gutzwiller ansatz for the Bose-Hubbard model \cite{fisher,jaksch}. Normalization dictates that $\sum_{n=0}^\infty |a_n|^2 = \sum_{n=0}^\infty |b_n|^2 = 1$. The average energy of our variational wavefunction is
\begin{equation}\label{eqn:Evar1}\begin{split}
 E_{\rm var} = &K\left( -zJ\left(\sum_n\sqrt{n}a_na_{n+1}\right)\left(\sum_n\sqrt{n}b_nb_{n+1}\right) \right. \\
 & +\sum_n \left(\frac{U}{4}n(n-1)-\frac{\mu}{2}n\right)(|a_n|^2+|b_n|^2)\\
 &\left. - \frac{U'}{4}\left(\sum_n n(|a_n|^2-|b_n|^2)\right)^2\right),
\end{split}\end{equation}
where $z$ is the number of nearest neighbors to a lattice site. In our case of a two-dimensional square lattice, $z=4$.

\subsection{Phase diagram}\label{subsec:phaseDiag1}

\subsubsection{$J=0$} \label{subsubsec:J=0homogeneous}

\begin{figure}[t]
\centering
\includegraphics[width=0.8\columnwidth]{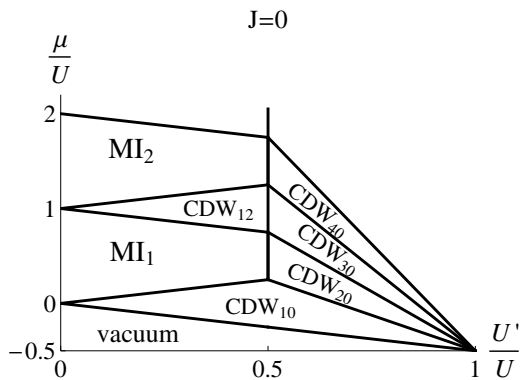}
\caption{Phase diagram of a homogeneous Bose gas in the absence of tunneling between lattice sites. The label $\rm{MI}_n$ denotes a Mott insulating phase with $n$ atoms on each lattice site, and $\rm{CDW}_{nm}$ denotes a charge-density wave phase with $n$ and $m$ atoms on even and odd sites, or vice versa. The parameters $U$ and $U'$ are the strengths of the short- and long-range interactions, while $\mu$ is the chemical potential.}
\label{fig:J=0}
\end{figure}

In the case of a deep optical lattice ($J=0$), the variational wavefunction which minimizes the energy describes an insulator with parameters
\begin{equation}\label{eqn:cdw}\begin{array}{rcl}
 a_n &=& \delta_{n,n_e},\\
 b_n &=& \delta_{n,n_o}.
\end{array}\end{equation}
Here, $\delta_{m,n}$ is the Kronecker $\delta$, and $n_e$ and $n_o$ are integers obtained by minimizing the energy in Eq. (\ref{eqn:Evar1}). If $n_e=n_o$, the ground state is a Mott insulator (MI). If $n_e\neq n_o$, the ground state is a charge-density wave insulator (CDW). The phase diagram for $J=0$ is plotted in Fig. \ref{fig:J=0}. All the phase transitions in this phase diagram are of first order.

We define the imbalance in the ground state to be
\begin{equation}\label{eqn:Idefn}
 I = \left|\frac{\sum_i (-1)^{x_i+y_i}\expect{\hc_i\+\hc_i}}{ \sum_i \expect{\hc_i\+\hc_i}}\right|.
\end{equation}
For $J=0$, the ground state is balanced ($I=0$) or partially imbalanced ($I<1$) for $\frac{U'}{U}<\frac{1}{2}$, and fully imbalanced ($I=1$) for $\frac{U'}{U}>\frac{1}{2}$. The wavefunction collapses to a state with infinite particles on every site for $U'>U$.

\subsubsection{$J\neq0,\ 0\leq U'\leq U/2$}

\begin{figure}
\begin{center}
\unitlength=0.74375in
\begin{tabular}{p{3em} p{0.8\columnwidth}}
\vspace{0.25\columnwidth} (a) & \vspace{0 pt}
\includegraphics[width=0.7\columnwidth]{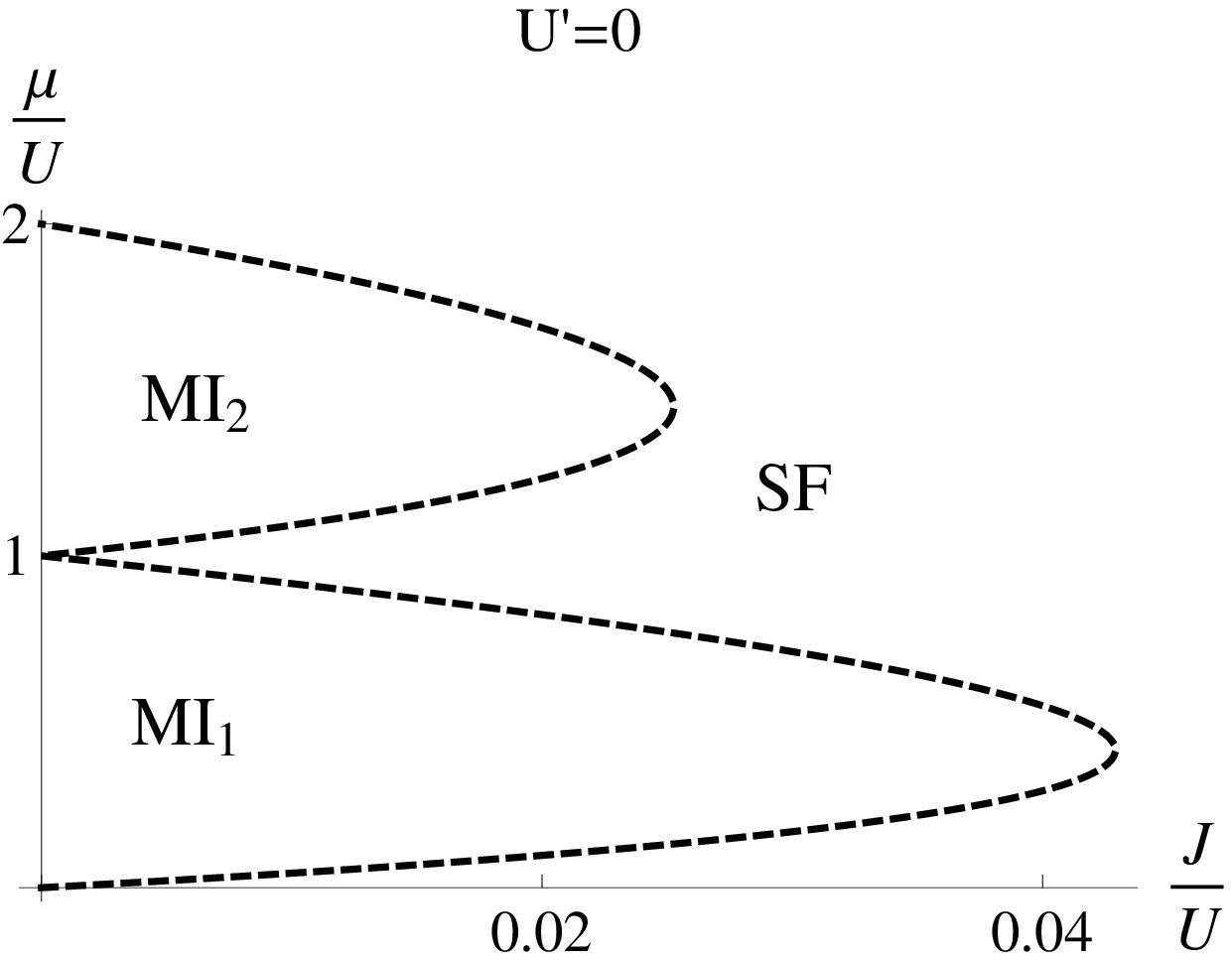}\\
\vspace{0.25\columnwidth} (b) & \vspace{0 pt}
\includegraphics[width=0.7\columnwidth]{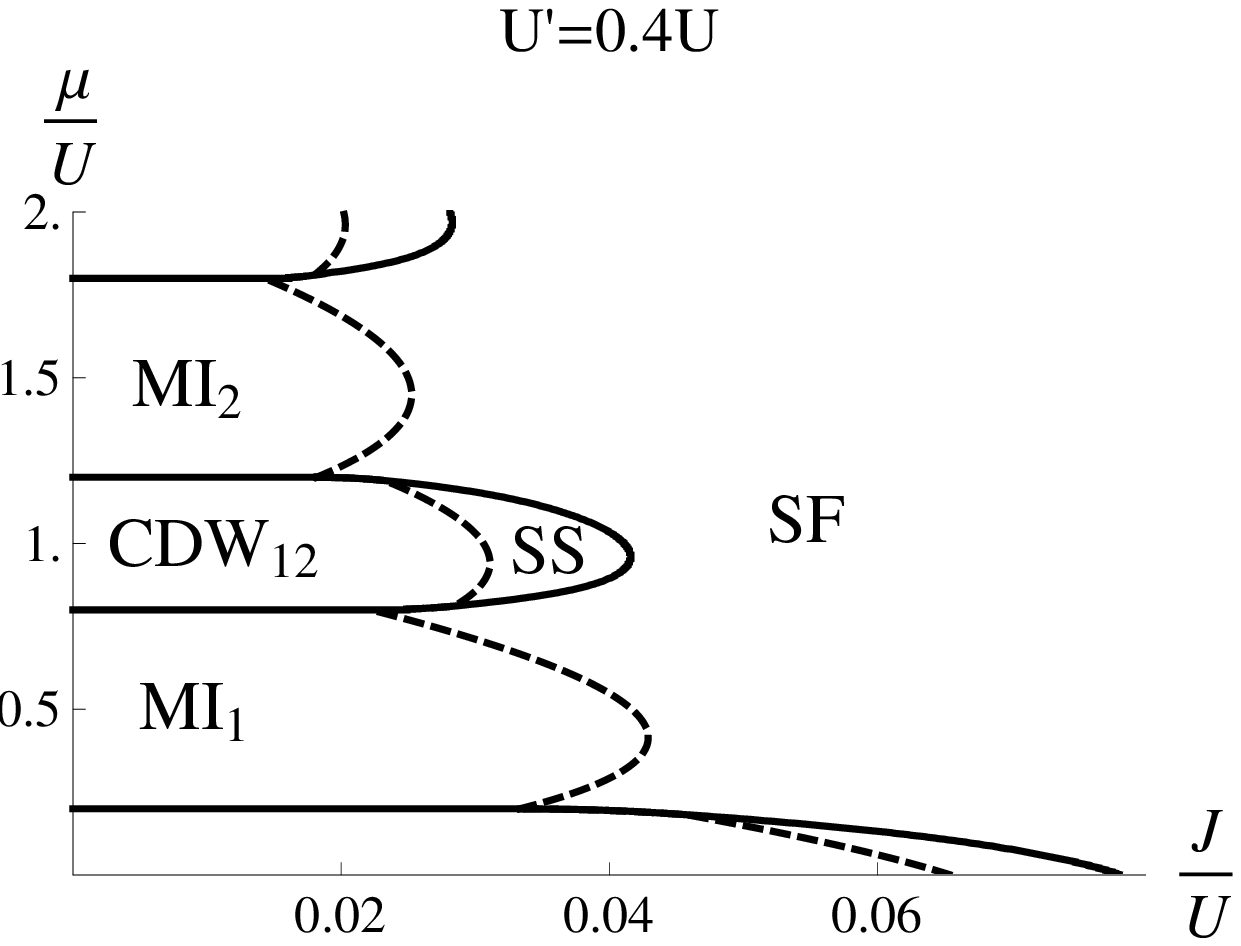}\\
\vspace{0.25\columnwidth} (c) & \vspace{0 pt}
\includegraphics[width=0.7\columnwidth]{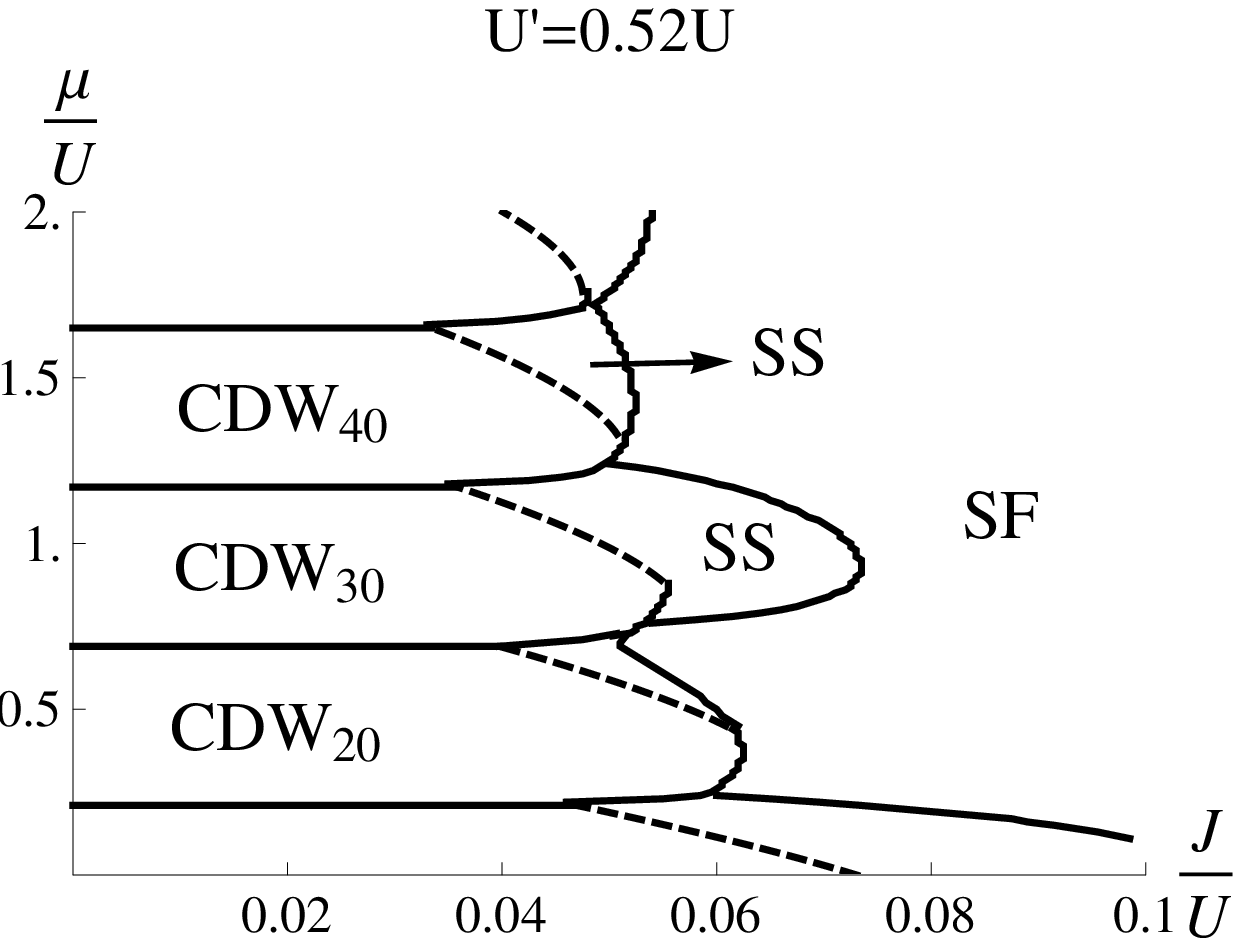}\\
\vspace{0.25\columnwidth} (d) & \vspace{0 pt}
\includegraphics[width=0.7\columnwidth]{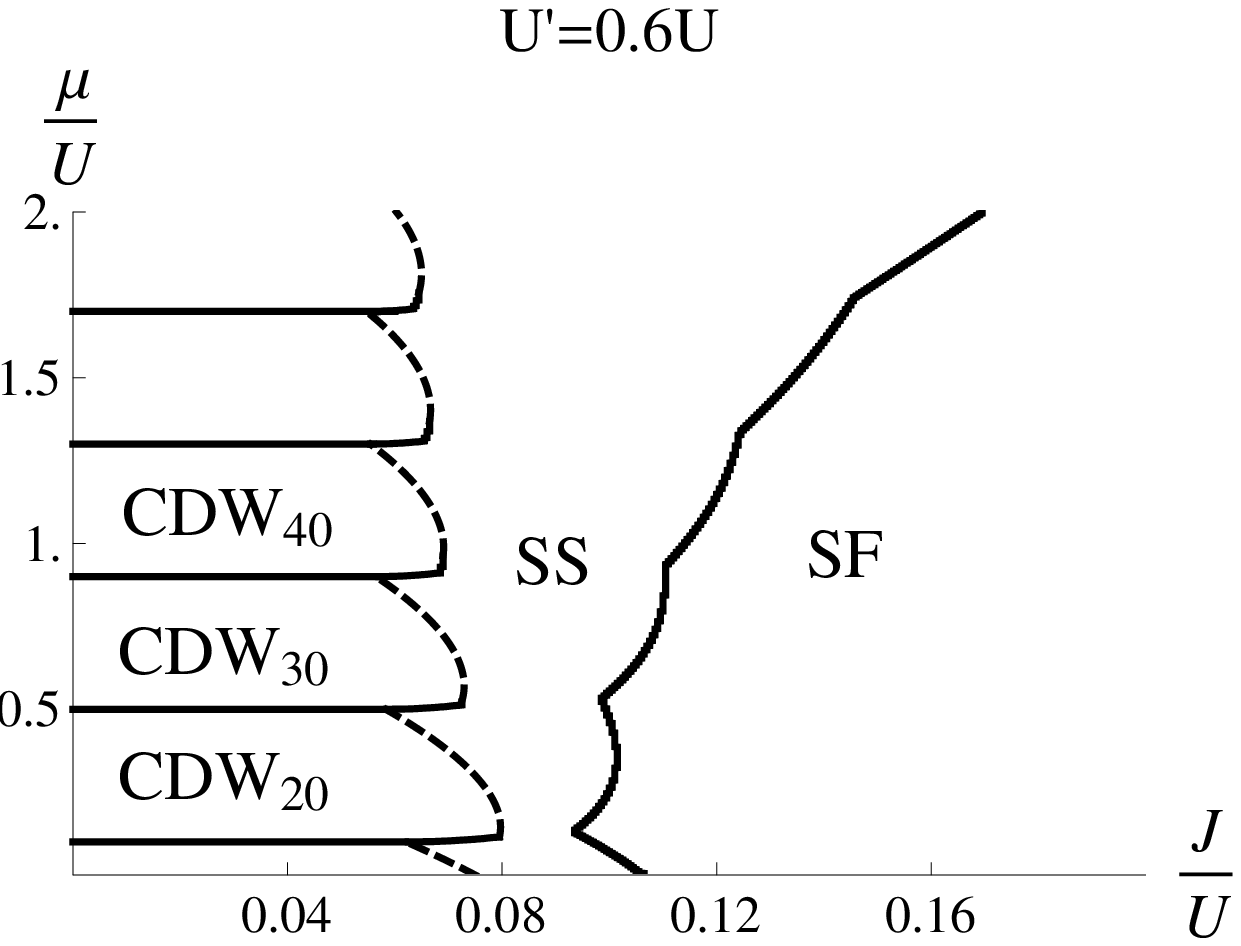}
\end{tabular}
\end{center}
\caption{Phase diagram of a homogeneous Bose gas for four different values of the long-range interaction strength $U'$. SS denotes supersolid, and SF denotes superfluid. Dashed lines indicate second-order phase transitions, and solid lines indicate first order. The parameter $J$ is the tunneling strength between lattice sites.}
\label{fig:Jneq0}
\end{figure}

For finite tunneling strength $J$, we minimize the energy in Eq. (\ref{eqn:Evar1}) numerically. The ground state is a superfluid (SF) if $\expect{c_i}$ is uniform and non zero. The ground state is a supersolid (SS) if $\expect{c_i}\neq\expect{c_j}\neq0$, where $i$ and $j$ are even and odd sites. Phase diagrams for four different values of $U'$ are plotted in Fig. \ref{fig:Jneq0}.

For $0<\frac{U'}{U}<\frac{1}{2}$, partially imbalanced CDW and SS phases (with $I<1$) appear in regions where $nU-\frac{U'}{2}<\mu<nU+\frac{U'}{2}$, where $n$ is any integer. In these phases, the density on even and odd sites is unequal, and the symmetry between even and odd sites is spontaneously broken. In most regions of the CDW lobes, the CDW phase undergoes a second-order phase transition to the SS phase as $J$ is increased at constant $\mu$ and $U'$. We define $\Omega$ to be the determinant of the Hessian of the free energy in Eq. (\ref{eqn:Evar1}), computed at the optimal variational parameters in Eq. (\ref{eqn:cdw}). At the second-order phase transition from SS to CDW, $\Omega=0$. This yields a simple analytic expression for the phase boundary,
\begin{equation}\label{eqn:Janal}
 J = \frac{1}{z}\sqrt{\frac{ \left(U'^2-(\mu-Un)^2\right)\left((U-U')^2-(\mu-Un)^2\right) }{ (\mu+U)^2-U'^2 }},
\end{equation}
where $n$ is the occupation number on lattice sites with fewer atoms, in the CDW phase at $J=0$. Our numerics confirm this result. Upon increasing $J$ further, the SS phase undergoes a first-order phase transition to the SF phase. Near the edges of the CDW lobes ($\mu\sim nU\pm\frac{U'}{2}$), the CDW phase directly undergoes a first-order phase transition to SF. There are no analytic expressions for these first-order phase boundaries.

\subsubsection{$J\neq0,\ U/2<U'< U$}

In this regime, the ground state is always a fully imbalanced ($I=1$) CDW phase at $J=0$. In this ground state, all the odd sites are empty, and the even sites have $n=\lceil\frac{\mu+U/2}{U-U'}\rceil$ atoms each, or vice versa. The ground state undergoes a first-order phase transition between different CDW phases periodically as $\mu$ is increased at constant $J$. For $U'$ near $U/2$, the CDW regions are partially surrounded by SS lobes [see Fig. \ref{fig:Jneq0}(c)]. As $U'$ is increased further, the SS lobes grow in size and connect together to form a continuous SS region [see Fig. \ref{fig:Jneq0}(d)]. In Figs. \ref{fig:Jneq0}(c) and \ref{fig:Jneq0}(d), the second-order transitions from CDW to SS are indicated by a dashed line. The phase boundaries of these second-order transitions are given by
\begin{equation}
 J = \frac{1}{z}\sqrt{\frac{ (U'n-\mu)\left((U-U')n-\mu\right)\left(\mu+U-(U-U')n\right) }{ U+\mu+U'n }}.
\end{equation}
The SS undergoes a first-order phase transition to SF as $J$ is increased further.

\section{Inhomogeneous gas}\label{sec:inhomogeneous}

In this section, we explore the phase diagram of a Bose gas in a harmonic trap, in the presence of infinite-range interactions mediated by an optical cavity. In experiments, the number of atoms $N$ can be a control parameter. Further, the total number of sites $K$ is not well defined. Thus, it is convenient to define $V = U'\frac{N}{K}$, and rewrite Eq. (\ref{eqn:Hcav}) as
\begin{equation}
 \hat{H}_{\rm cav} = -\frac{V}{N}\left(\sum_i (-1)^{x_i+y_i}\hc_i\+\hc_i\right)^2.
\end{equation}
To find the ground state of this model, we generalize Eq. (\ref{eqn:Gutzwiller1}), writing
\begin{equation}
 \ket{\psi} = \sum_i \sum_{n=0}^\infty \frac{a_{ni}}{\sqrt{n!}}\left(\hc_i\+\right)^n\ket{0}.
\end{equation}
The variational energy is then
\begin{equation}\label{eqn:Evar2}\begin{split}
 E_{\rm var} = & -J\sum_{\langle ij\rangle} \left(\sum_n\sqrt{n}a_{ni}a_{n+1,i}\right)\left(\sum_n\sqrt{n}a_{nj}a_{n+1,j}\right)  \\
 & +\sum_{n,i} \left(U\frac{n(n-1)}{2}-\mu_i n\right)|a_{ni}|^2\\
 & - V\frac{ \left(\sum_{n,i} (-1)^{x_i+y_i}n|a_{ni}|^2\right)^2 }{ \sum_{n,i}n|a_{ni}|^2 }.
\end{split}\end{equation}

We minimize $E_{\rm var}$ with respect to all the variational parameters. Due to the presence of infinite-range interactions in our model, traditional methods of treating spatially varying potentials, such as the local-density approximation (LDA), fail. We demonstrate the failure of the local-density approximation in the insulating phases in Sec. \ref{subsec:J=0}. We present our numerical results for the phase diagram in Sec. \ref{subsec:phaseDiag2}

\subsection{J=0}\label{subsec:J=0}

\begin{figure}[t]
\unitlength=0.6in
\centering
\begin{tabular}{cc}
\includegraphics[width=0.35\columnwidth]{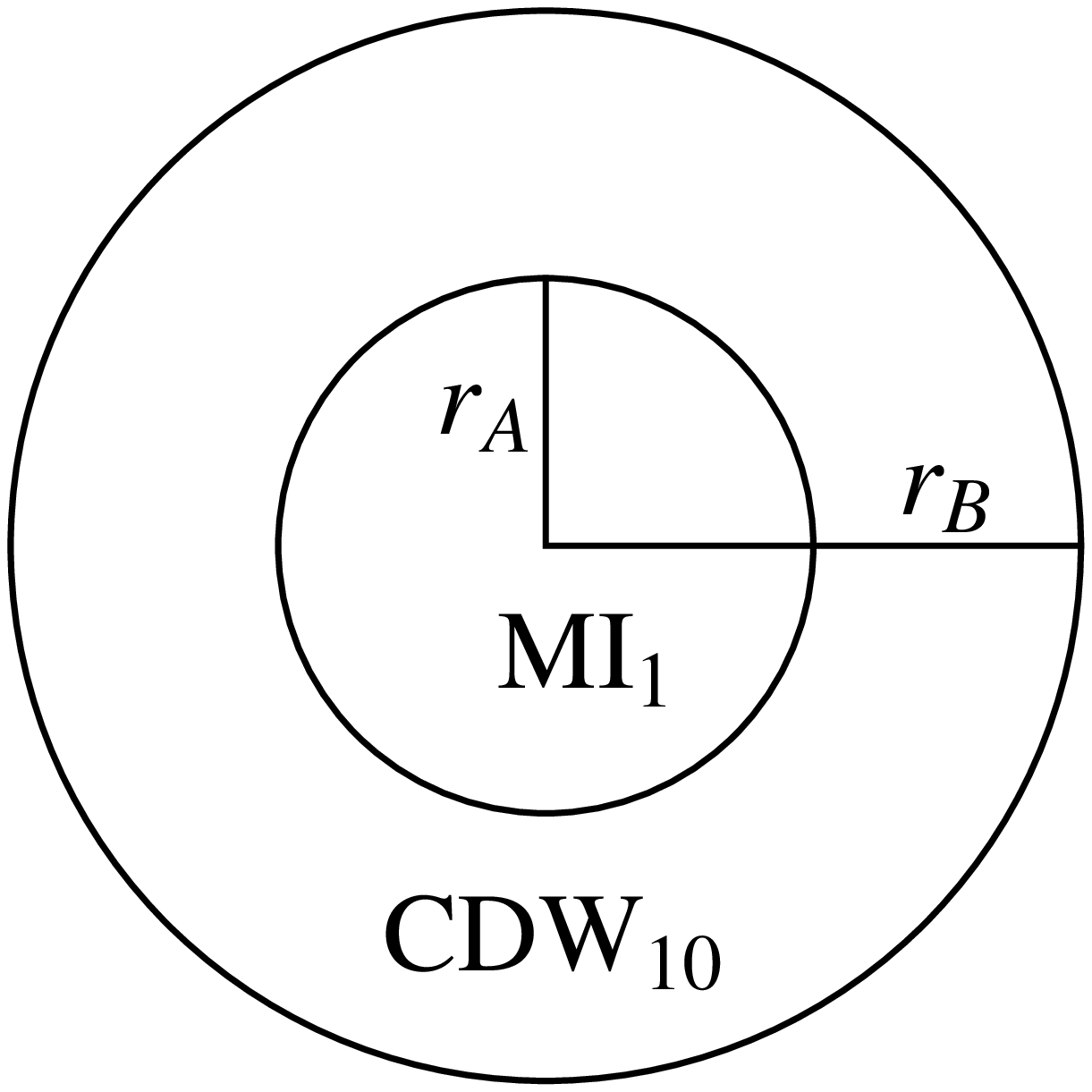} & \includegraphics[width=0.6\columnwidth]{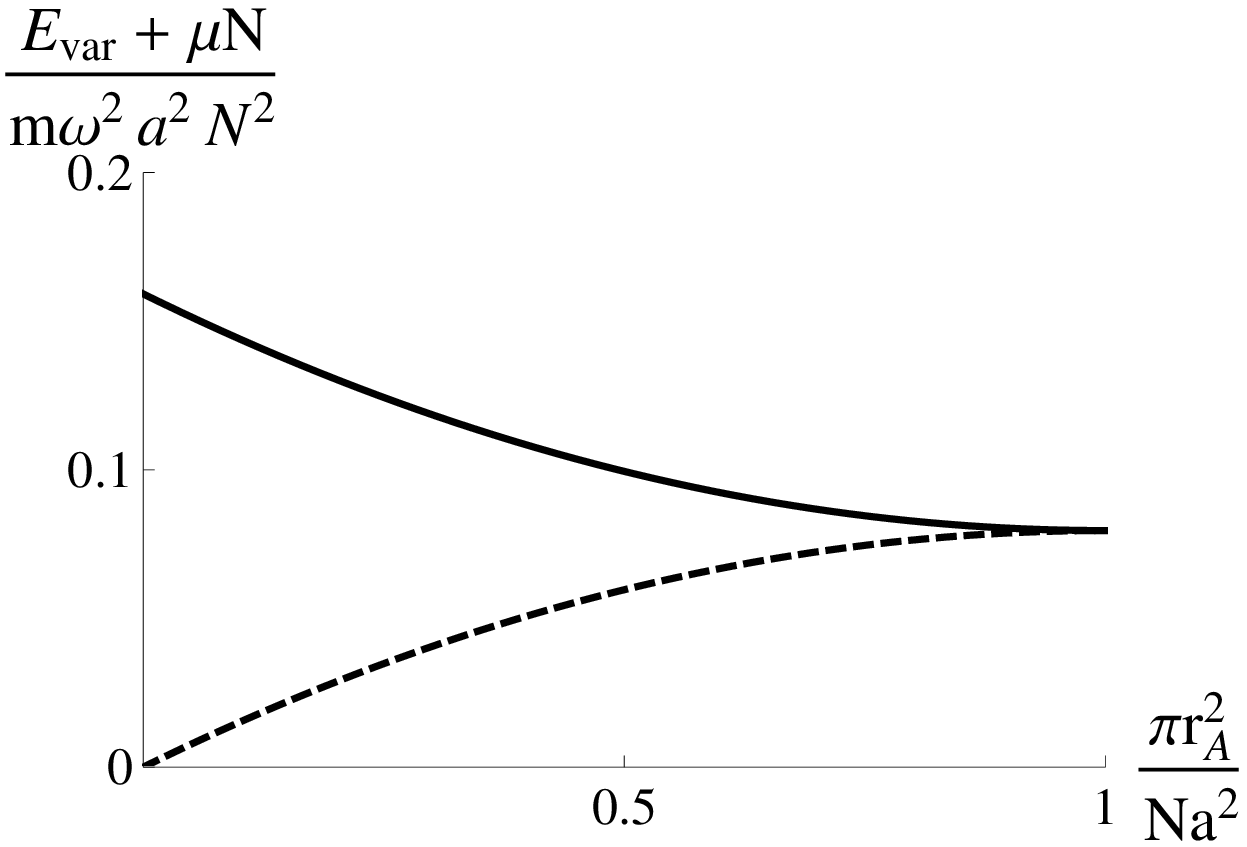}\\
 & \\
(a) & (b)
\end{tabular}
\caption{(a) A $n=1$ Mott insulating core of radius $r_A$, surrounded by a CDW$_{10}$ ring from radius $r_A$ to $r_B$. (b) Scaled energy $E_{\rm var}$ as a function of scaled radius $r_A$ for two cases: $V=0$ (solid) and $V=\frac{Nm\omega^2a^2}{2\pi}$ (dotted).}
\label{fig:MI,CDW}
\end{figure}
\begin{figure}[t]
\begin{center}
\includegraphics[width=0.7\columnwidth]{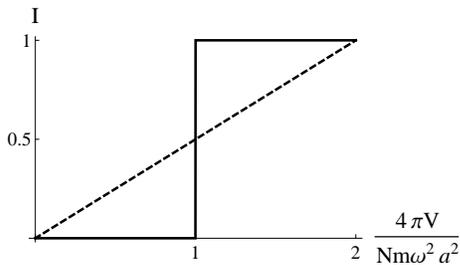}
\end{center}
\caption{Comparison of LDA and full theory. Vertical axis shows the particle imbalance $I$ between even and odd sites. Horizontal axis shows the strength of the long-range interaction. LDA (dashed line) shows a continuous growth of the imbalance, while the full theory (solid line) shows a discontinuity.}
\label{fig:imbalance}
\end{figure}

In the absence of tunneling, every lattice site has an integer number of atoms. For most of the experimentally relevant parameters, it suffices to consider only zero or one atom on every site. One expects the cloud to have a MI$_1$ core of radius $r_A$ with unit filling, as shown in Fig. \ref{fig:MI,CDW}(a). This core is expected to be surrounded by a CDW$_{10}$ ring extending from radius $r_A$ to radius $r_B$. The CDW$_{10}$ ring is surrounded by vacuum. In the limit of a slowly varying trap, the variational energy can be approximated as
\begin{equation}\begin{split}
 E_{\rm var} &= \frac{1}{a^2}\int_0^{r_A}2\pi rdr\left(\frac{1}{2}m\omega^2r^2-\mu\right)\\
 &+ \frac{1}{a^2}\int_{r_A}^{r_B}\pi rdr\left(\frac{1}{2}m\omega^2r^2-\mu\right) -
 \frac{V}{N}\left(\frac{\pi(r_B^2-r_A^2)}{2a^2}\right)^2\\
 = &\frac{\pi m\omega^2}{8a^2}(r_A^4+r_B^4) - \frac{\pi\mu}{2a^2}(r_A^2+r_B^2) -
 \frac{V}{N}\left(\frac{\pi(r_B^2-r_A^2)}{2a^2}\right)^2.
\end{split}\end{equation}
Fixing the number of particles $N = \frac{\pi}{2a^2}(r_A^2+r_B^2)$, the variational energy is
\begin{equation}\label{eqn:EvarHarmonic}
 E_{\rm var} = \left(\frac{m\omega^2a^2}{4\pi}-\frac{V}{N}\right)\left(N-\frac{\pi r_A^2}{a^2}\right)^2 + \frac{N^2m\omega^2a^2}{4\pi}-\mu N.
\end{equation}
This variational energy is plotted as a function of $r_A$ in Fig. \ref{fig:MI,CDW}(b). The energy minimum occurs at
\begin{equation}
 r_A = \left\{ \begin{array}{lcr}
 0 & \mbox{if} & V>\frac{Nm\omega^2a^2}{4\pi},\\
 \sqrt{\frac{Na^2}{\pi}} = r_B & \mbox{if} & V<\frac{Nm\omega^2a^2}{4\pi}.
 \end{array}\right.
\end{equation}
The ground state transitions from a completely Mott insulating gas to a completely checkerboarded gas at the critical value $V=\frac{Nm\omega^2a^2}{4\pi}$, where the imbalance between even and odd sites undergoes an abrupt jump from $0$ to $1$ (see solid line in Fig. \ref{fig:imbalance}). We contrast this result to the predictions of LDA (which are not valid because of the long-range interactions). In traditional LDA, the local phase at position $\vec{r}$ is that of a homogeneous system with chemical potential $\mu(\vec{r})$. As seen in Fig. \ref{fig:J=0}, this implies that unless $V=0$, one always has a CDW ring. A straightforward calculation of the imbalance between even and odd sites shows that within this approximation, the imbalance grows gradually as $V$ is increased. The dashed line in Fig. \ref{fig:imbalance} depicts the imbalance obtained from this LDA calculation, with $K=2N$ lattice sites.

\subsection{Phase diagram}\label{subsec:phaseDiag2}

\begin{figure}[t]
\unitlength=0.64in
\begin{center}
\begin{tabular}{p{2em} p{0.98\columnwidth}}
\vspace{0.25\columnwidth} (a)&
\vspace{0 pt}\hspace{10 pt}\includegraphics[width=0.8\columnwidth]{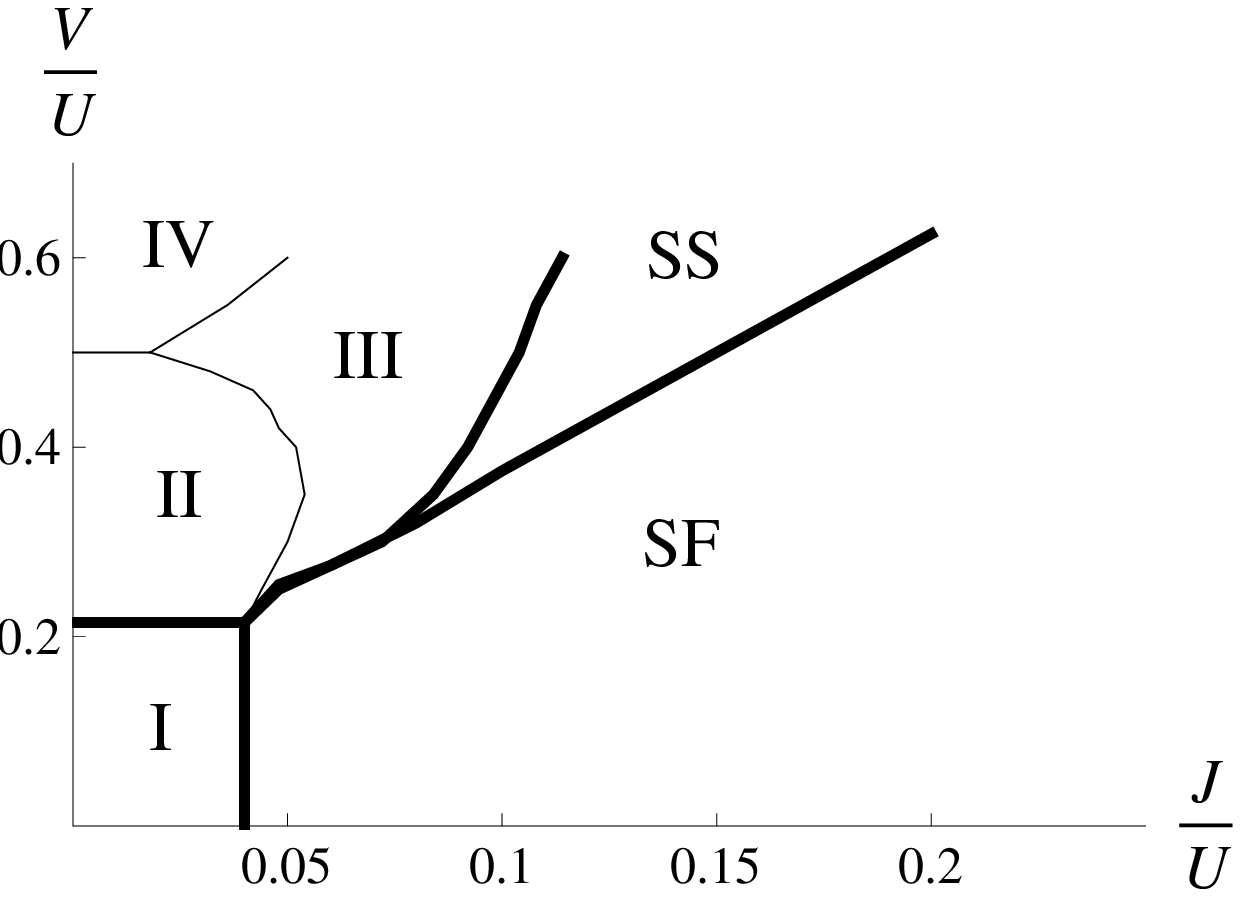} \\
\vspace{0.25\columnwidth} (b)&
\vspace{0 pt}\hspace{10pt}\includegraphics[width=0.8\columnwidth]{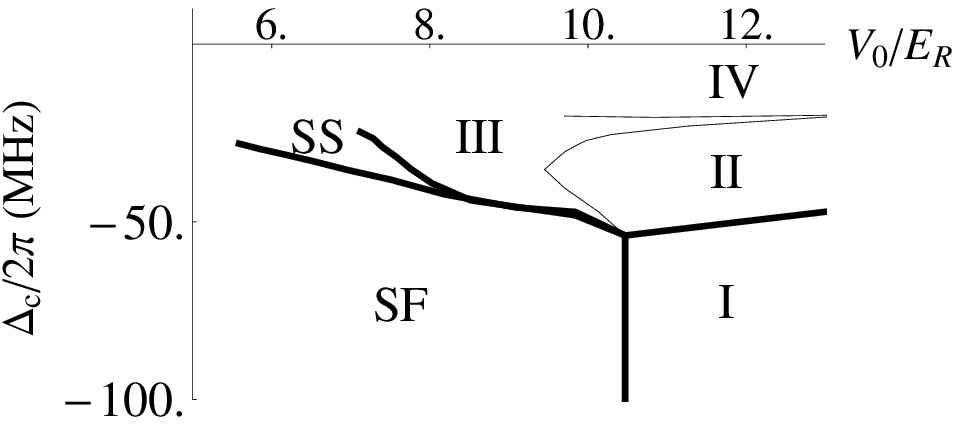}
\end{tabular}
\end{center}
\caption{Phase diagram for an inhomogeneous Bose gas in a harmonic trap and an optical lattice, plotted against (a) model parameters, and (b) experimentally relevant parameters. Region I denotes MI$_1$-SF phase, II is CDW$_{10}$-SS, III is SS-CDW$_{10}$-SS, and IV is CDW$_{20}$-SS-CDW$_{10}$-SS. Thick lines denote phase boundaries measured in experiments \cite{Esslinger}. The chemical potential is chosen to be $\mu=0.5U$, and the trap frequency is such that $m\omega^2a^2 = 0.01U$. In (b), $V_0$ is the lattice depth, and $\Delta_c$ is the detuning of the pump laser from the fundamental cavity mode. The Rabi frequency is chosen to be $\eta = \sqrt{\frac{2\pi}{N}}\frac{E_R}{\hbar}$, the scattering length is $100\ a_0$, and the lattice constant and lattice depth in the $z$ direction are $670\ \rm{nm}$ and $25 E_R$, as consistent with experiments \cite{Esslinger}.}
\label{fig:harmonicPhaseDiag}
\end{figure}

In this section, we use numerical methods to calculate the phase diagram of an inhomogeneous Bose gas in a harmonic trap. We work in a relatively small density regime, and truncate the ansatz to allow $0$, $1$, or $2$ atoms per site. For a $35\times35$ square lattice, we numerically minimize the variational energy in Eq. (\ref{eqn:Evar2}) with respect to $2450$ independent variational parameters. In the ground state, the atoms arrange in concentric shells of insulating ($\rm{MI}_n$/$\rm{CDW}_{mn}$) and conducting (SS/SF) regions. We label the state of the gas at every point in the phase diagram by listing the phases of the atoms in these shells, in the order that they occur outwards from the center of the cloud. For example, SF denotes that the entire cloud is superfluid, and MI$_1$-SF denotes that the center of the cloud is in the Mott insulating phase with unit filling, surrounded by a superfluid ring. In Fig. \ref{fig:harmonicPhaseDiag}, we show the phase diagram for $\mu=0.5U$ and $m\omega^2a^2=0.01U$. For the range of hopping and long-range interaction that we consider, we find six different ways that the atoms arrange, namely SF, MI$_1$-SF, SS, SS-CDW$_{10}$-SS, CDW$_{10}$-SS, and CDW$_{20}$-SS-CDW$_{10}$-SS. We find similar results for other parameters, but the exact locations of the phase boundaries differ. Below, we discuss some of the interesting features in the phase diagram in Fig. \ref{fig:harmonicPhaseDiag}.

The MI$_1$-SF and SF phases have no sublattice imbalance, $I=0$. In these phases, the long-range interaction term involving $V$ does not contribute to the free energy. Therefore in this regime, the phase boundary between SF and MI$_1$-SF does not depend on $V$, and the boundary is a vertical line at $J = \frac{\mu(U-\mu)}{z(U+\mu)}\simeq0.04U$ for our parameters.

In the absence of tunneling and for a fixed chemical potential $\mu$, the size of the cloud in the MI$_1$-SF phase is $r = \sqrt{\frac{2\mu}{m\omega^2}}$. The free energy [from Eq. (\ref{eqn:EvarHarmonic})] is $E_{\rm var} = -\frac{\pi\mu^2}{m\omega^2a^2}$. In the CDW$_{10}$-SS phase, the cloud extends up to $r = \sqrt{\frac{2(\mu+V)}{m\omega^2}}$. The free energy of the CDW$_{10}$-SS phase is $E_{\rm var} = -\frac{\pi(\mu+V)^2}{2m\omega^2a^2}$. By comparing the free energies in the two phases, we find that the phase boundary between MI$_1$-SF and CDW$_{10}$-SS approaches $V = \left(\sqrt{2}-1\right)\mu\simeq0.2U$ as $J\rightarrow0$. Numerically, we find that this is a good approximation even for $J\neq0$.

A CDW$_{20}$ core appears inside the CDW$_{10}$ region if it is energetically cheaper to add atoms to the center rather than the edge of the cloud. In the absence of tunneling, the energy cost of adding an atom to an occupied site in the center of a CDW$_{10}$ cloud is $U-\mu-V$. There is no energy cost for adding an atom at the edge of a CDW$_{10}$ cloud. Therefore in the absence of tunneling, the phase boundary between CDW$_{10}$-SS and CDW$_{20}$-SS-CDW$_{10}$-SS occurs at $V = U-\mu=0.5U$. Numerically, we find that this is a good approximation even for $J\neq0$.

The phase diagram in Fig. \ref{fig:harmonicPhaseDiag} exhibits a multicritical point at $J\sim0.04U, V\sim0.2U$. At this multicritical point, the SF, MI$_1$-SF, CDW$_{10}$-SS, and SS-CDW$_{10}$-SS phases coexist. We find two other tricritical points at $J\sim0.06U, V\sim0.27U$, and $J\sim0.018U, V\sim0.5U$.

\subsection{Comparison to experiment}

Experimentalists in Zurich attempted to generate a phase diagram similar to Fig. \ref{fig:harmonicPhaseDiag} \cite{Esslinger}. By monitoring the intensity in the cavity, they could detect a transition from a state with no sublattice imbalance to one in which imbalance is present. For example, this technique can find the transition from SF to SS. The researchers also monitored the condensate fraction as a function of lattice depth, finding kinks which they interpreted as phase transitions. Indeed, the appearance of an insulating region should generate such a kink. In generating their figures, the researchers only include the kink at largest $J/U$. The resulting phase diagram agrees well with the thick lines in Fig. \ref{fig:harmonicPhaseDiag}(b). Further analysis of their data should reveal the other curves in Fig. \ref{fig:harmonicPhaseDiag}(b).

\section{Summary}\label{sec:conclusions}
We calculated the phase diagram of a two-dimensional Bose gas with short-range and long-range checkerboard interactions in an optical lattice. The long-range checkerboard interactions are produced by trapping the Bose gas in a single mode Fabry-Perot cavity, and illuminating it with a laser beam in the transverse direction. We found that, in the presence of these interactions, the Bose gas exhibits four phases - a Mott insulator with integer filling, a charge-density wave with different integer fillings on even and odd sites of the lattice, a superfluid with off-diagonal long-range order, and a supersolid with SF and CDW orders. We presented numerical results for the phase diagram of this homogeneous gas, and obtained analytical expressions for all the second-order phase boundaries. We also presented numerical results for the phase diagram of an inhomogeneous gas in a harmonic trap. Our numerical phase diagram agrees well with the phase diagram that was experimentally measured recently \cite{Esslinger}. We predict that further analysis will reveal more phases in their data.

The system considered in this study is interesting for several reasons. First, due to long-range interactions, LDA fails. Second, the experiments in \cite{Esslinger} made the first detection of a supersolid phase. One caveat is that this supersolid phase is a bit unusual, as it breaks only a global symmetry and not a local symmetry. Thus one would never expect to see domain walls in the checkerboard order, unless they are imposed by using a cavity mode with nodes \cite{benlev}. Beyond this system, atoms coupled to optical cavities provide an avenue to control atom-atom interactions. Similar setups which trap atoms in multimode cavities could be used to produce controllable medium-range interactions between atoms \cite{benlev}. Multimode cavities can also be used to create phononlike excitations in the lattice. Atomic clouds trapped in cavities can be used to explore non-equilibrium phases \cite{polariton} and nontrivial phase transitions in driven dissipative quantum systems \cite{noneq1,noneq2,noneq3}.

\section*{ACKNOWLEDGMENT}
This material was based upon work supported by the National Science Foundation under Grant No. PHY-1508300.

\bibliography{references}
\end{document}